# THz near-field nanoscopy employing coherent synchrotron radiation

Alexander Veber[1,2], Maria Eleonora Temperini[1], Ljiljana Puskar[1], Ulrich Schade[1]

*[1] Institute for Electronic Structure Dynamics, Helmholtz-Zentrum Berlin für Materialien und Energie GmbH, Albert-Einstein-Str. 15, 12489 Berlin, Germany*

*[2] Department of Chemistry, Humboldt-Universität zu Berlin, Brook-Taylor-Straße 2, 12489 Berlin, Germany*

**Abstract**

We demonstrate that the coherent synchrotron radiation (CSR) can be used for the scattering type near-field scanning optical microscopy (s-SNOM). Using a commercially available s-SNOM coupled to a synchrotron source we perform THz imaging and spectroscopy experiments in the spectral range of 25-55 $cm^{-1}$ and spatial resolution of 60 nm. It is shown that the same optical configuration can be used for nanoscopy in the range of 25 to 4000 $cm^{-1}$. Conducting experiments across the THz, far-, and mid-IR spectral ranges within a single setup offers a unique capability for the ultrabroadband characterization of complex materials in both life and material sciences at the nanoscale.

**Introduction**

In the last years the infrared (IR) scattering type near-field scanning optical microscopy (s-SNOM) became a versatile and widely used analytical tool. The method enables detailed chemical characterization by means of IR spectroscopy far beyond the diffraction limit at the nanometer spatial scales. In the s-SNOM methodology, the spatial resolution is determined by the metallized AFM probe and is therefore independent of the light wavelength, allowing to maintain nanoscale resolution across spectral ranges from the visible to the THz.[1,2]

Significant efforts were aimed towards implementation of the IR s-SNOM technique in the mid-IR that allows identification of the molecules and materials from the gained spectral information. Today modern tunable IR laser sources are capable of accessing the spectral range of approximately 550–7000 $cm^{-1}$, covering the most demanded fingerprint spectral region.[3]

Alongside table-top systems, the IR s-SNOM technique has been implemented at large-scale facilities, i.e. synchrotron IR beamlines[4,5] and free-electron lasers.[6] Implementation of the s-SNOM at FELBE demonstrated that this approach enables imaging and point-by-point spectroscopy in the spectral range of 40-2000 $cm^{-1}$.[7,8] The use of the synchrotron radiation remains beneficial for the near-field broadband infrared spectroscopy due to its brilliance over extremely broad spectral range and unprecedented stability.[4,5,9] The spectral range available at the synchrotron is limited by the availability of the fast low-noise optical detectors and the light flux in a specific spectral region, which depends on the parameters of the storage ring and the beam-line optical design.

Extending nano-spectroscopy in the lower energy range allows to probe phenomena such as collective lattice vibrations, plasmonic oscillations, charge carriers and intermolecular interactions.[10] THz time-domain systems enable measurements in the range of 0.5–2 THz (15-70 $cm^{-1}$),[11] however the time-domain implementation makes this system incompatible with standard s-SNOM systems based on an asymmetric interferometer design.

Recent advancements at synchrotron facilities have implemented s-SNOM setups with a liquid He-cooled MCT detector enabling ultrabroadband near-field spectroscopy down to 160 $cm^{-1}$,[12] while the use of a Ge:Ga detector further extended the accessible spectral range to 100 $cm^{-1}$.[13,14] Nevertheless, the synchrotron-based THz spectroscopy in the range < 100 $cm^{-1}$ is challenging due to the natural significant drop in dipole magnet radiation intensity at lower energies during common multi-bunch operational mode of a storage ring.[15]

Significantly higher THz emission is observed in the low-alpha operation mode of a storage ring. Compressed bunches in the low-alpha mode naturally produce coherent synchrotron radiation (CSR), i.e. its intensity scales as $N^2$, where N is number of electrons, and its maximum lays in the hundreds of GHz to few THz spectral range.[16] It has been demonstrated that CSR at BESSY II storage ring can be used for the near-field THz imaging: the spatial resolution of ≥ 70 µm, determined by the used conical aperture probes, enabled imaging beyond the diffraction limit at the wavelength λ≈0.83 mm (~12 $cm^{-1}$).[17]

In this work we demonstrate that the low-alpha operation mode of the BESSY II storage ring can be used for the s-SNOM based THz near-field spectroscopy and imaging. The use of CSR allows to extend the spectral range for the nano-spectroscopy and imaging down to 25 $cm^{-1}$ (400 µm), that in combination with the multi-bunch operation mode allows to conduct the experiments from 25 to 4000 $cm^{-1}$.

**Experimental setup**

IR nano spectroscopy and imaging experiments were done at the IR-nanospectroscopy end-station of IRIS beam-line at BESSY II synchrotron,[18] using a s-SNOM setup (neaScope, Attocube, Germany) coupled to the brilliant broadband synchrotron infrared light.[19] The measurements in the THz spectral range (<100 $cm^{-1}$) were done during low-alpha operation mode of the BESSY II storage ring. The complementary near-field spectra in the range of 350-4000 $cm^{-1}$ were collected in the multi-bunch operational mode of the electron storage ring.

To make possible the THz experiments a diamond wedged window was installed at the corresponding port of the beamline and a 2.2 mm thick, high-resistivity float zone silicon plane parallel window was used as a beam-splitter inside the s-SNOM. An InSb hot electron bolometer (Model QFI/2BI, QMC Instruments Ltd, UK) was used as the detector. The detector offers both sensitive and fast detection of the radiation with energies below 60 $cm^{-1}$, having the maximum of the sensitivity at approximately 30 $cm^{-1}$ (Figure S1), and bandwidth of approximately 1 MHz. The incoming higher energy radiation is blocked by a pair of long pass multi-mesh filters (cut-off edge is at 100 $cm^{-1}$) installed at 77 K and 4.2 K. In order to further suppress the incoming background radiation, we have installed a circular aperture of approximately 8 mm at 77 K in front of the f/2 Winston cone of the detector (see Figure S2 for the details). This allows to match the detector optical étendue to that of the source. The benefits of such a modification to an IR detector for synchrotron-based IR spectroscopy experiments have been demonstrated previously.[20] Gold-coated AFM probes with typical tip radius < 50 nm, nominal spring constant of 7.4 N/m and a resonance oscillation frequency of 160 kHz (PPP-NCSTAu, Nanosensors™, Switzerland) were used for the THz experiments.

A photoconductive LHe-cooled Si:B detector(Infrared Laboratories, USA) and a pellicle beam-splitter (BP145B4, Thorlabs, Germany) were used for the experiments in the spectral range of 350-800 $cm^{-1}$. A $LN_2$-cooled photoconductive MCT detector (MCT-13-0.05, Infrared Associates, USA) and a ZnSe beam splitter (Attocube, Germany) were used for the spectral range of 800-4000 $cm^{-1}$. The Si:B detector was modified similarly to the InSb HEB, and an aperture of 1 mm was installed in the baffle of the detector at 77 K before the f/7.5 Winston cone. AFM-probes with typical tip radius <25 nm, nominal spring constant of 42 N/m, and a resonance oscillation frequency of approximately 255 kHz (Arrow™ NCPt, NanoWorld™, Switzerland) were used for the far- and mid-IR experiments.

**s-SNOM spectra modelling**

The third-harmonic demodulation of the SNOM amplitude and phase signals on $SiO_2$ was simulated using the finite dipole model (FDM) implemented in the Python package "snompy". [21]

The sample was modelled as a $SiO_2$ layer with a thickness of 18 nm on a semi-infinite Si substrate. The $SiO_2$ permittivity in the 350-1600 cm-1 spectral range was taken from the tabulated refractive index[22] available in the refractive index database.[23] The permittivity away from the phonon modes was assumed frequency-independent in the entire simulated range and fixed to 11.7, including the THz range (10-70 $cm^{-1}$). To reproduce the experimental conditions, for the mid- and far-IR simulations, the AFM tip was represented as an ellipsoid with a radius of 25 nm and a semi-major axis of 100 nm, oscillating with the amplitude of 70 nm. In the THz range, the AFM tip was modelled with a radius of 50 nm and a semi-major axis of 200 nm and the tapping amplitude of 80 nm. The angle of incidence of the light beam was fixed to 60°. The near-field response of the sample was referenced to that of Si in order to directly compare with the experimental demodulated signal.

## Results and Discussion

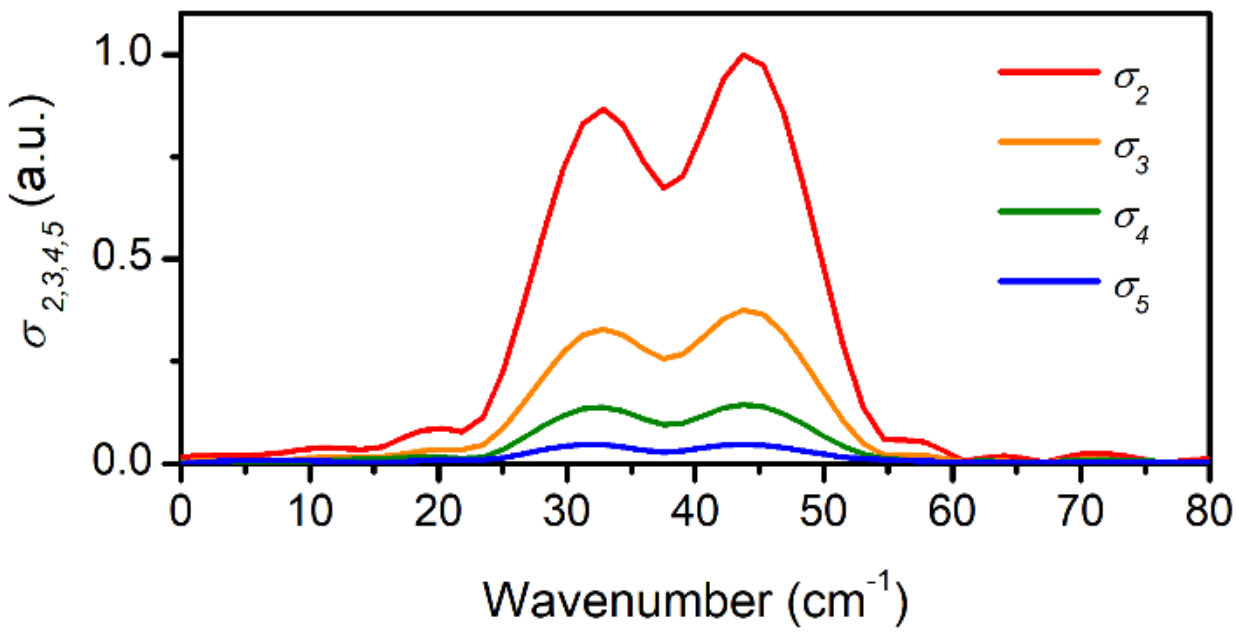

*Figure 1. Near-field optical amplitude spectrum at different harmonic orders collected on a gold surface using the CSR*

The optical near-field response in the low-alpha mode was tested on a gold-coated surface. The use of CSR enables spectroscopy in the spectral range of about 25 to 50 $cm^{-1}$ and the signal could be detected up to the $5^{th}$ demodulation harmonics, ensuring high suppression of the background signal (Figure 1). Previously it has been shown that the near-field response using CSR at BESSY II storage ring could be observed at the energies down to 12 $cm^{-1}$. According to the far-field measurement the CSR currently implemented at BESSY II storage ring spans from ca. 10 to 60 $cm^{-1}$ with a maximum at ~20 $cm^{-1}$, whereas the detector responsivity is the highest in the range < 40 $cm^{-1}$ (Figure S1). The significant increase of the low-energy limit observed in the s-SNOM configuration can be explained by the optical design of the commercial microscope, that is not optimized for the operation with the relatively large THz optical beam, and the near-field enhancement of the commercial AFM probe used in the experiment. It has been shown that THz can be significantly enhanced with use of longer metallized or solid metal probes.[24,25] Considering the low-alpha radiation spectrum and the spectral response of the detector (Figure S1) we expect that the spectral range available for the THz s-SNOM experiment can be extended down to 10 $cm^{-1}$ with the use of an AFM-probe designed for operation in the THz region and optimized optical outline of the microscope.

Complementary experiments utilizing the three available detectors enable near-field spectroscopy across a spectral range from 25 to 3900 $cm^{-1}$ (Figure 2). The current spectral gap of 50 to 350 $cm^{-1}$ could be significantly narrowed to 50–150 $cm^{-1}$ by coupling a LHe-cooled MCT detector to the microscope[12]. The spectrally seam-less spectroscopy could be achieved with use of a fast NbN superconducting hot electron bolometers.[26] Relevant development works at the IRIS-beamline BESSY II storage ring are currently in progress.

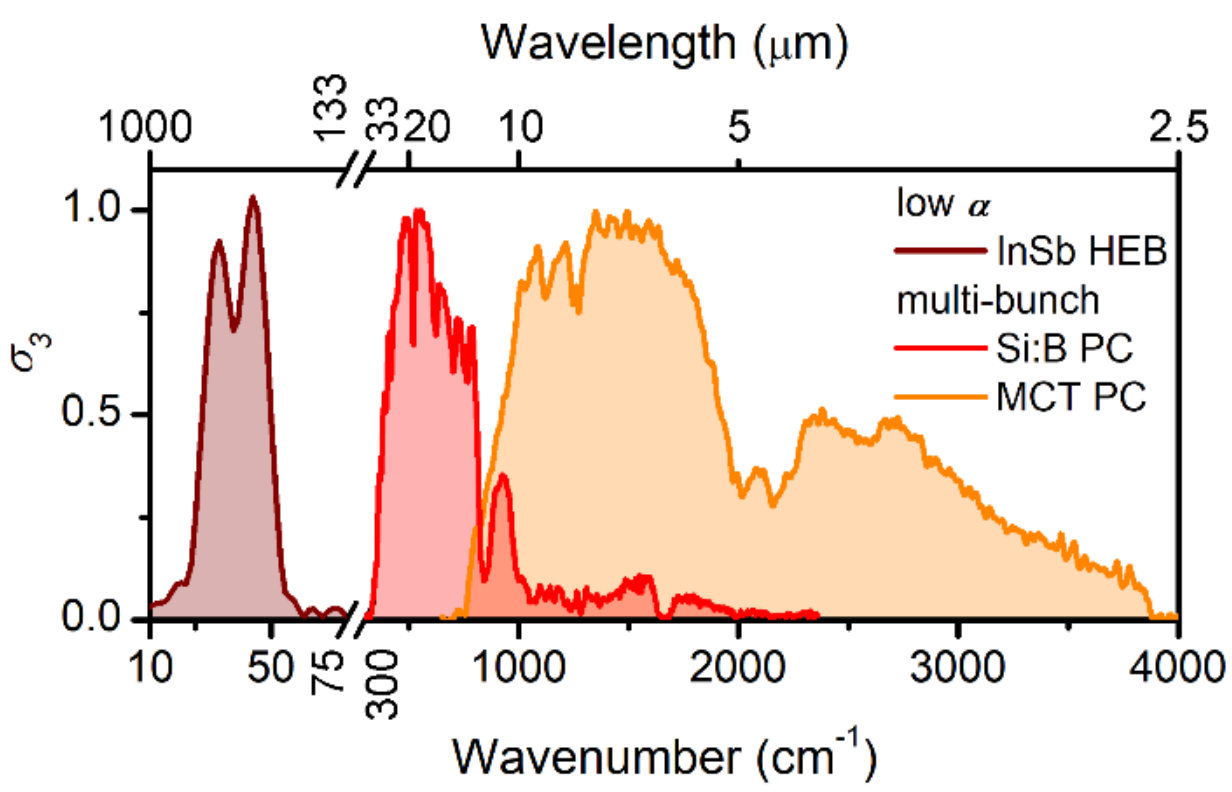

*Figure 2. Near-field optical amplitude spectrum collected on a gold surface. The THz spectrum (10-75 $cm^{-1}$) is measured using CSR and the InSb HEB detector; far-IR (300-1000 $cm^{-1}$) and mid-IR (750-4000 $cm^{-1}$) spectra recorded in multi-bunch operation mode of the synchrotron with use of Si:B and MCT detectors, respectively.*

An exemplary broadband near-field investigation with use of the three spectral ranges available was made for a $SiO_2$/Si composite sample. White-light optical images collected on the TGC-1 grating sample (TipsNano, Estonia) reveal the contrast between Si and $SiO_2$ (Figure 3A). Both the materials are transparent in the THz spectral range and the observed contrast of about 1.2 between two materials, calculated as the ratio of the optical amplitude signal at Si and $SiO_2$ is due to difference in the real part of the dielectric permittivity in this spectral range and match the value predicted by the FDM calculations (gray dashed trace in Figure 3C). As it can be seen from the linear profile perpendicular to the Si-$SiO_2$ border region (Figure 3B), the spatial resolution achieved in the experiment was ~60 nm, that is comparable to the nominal tip radius of the AFM probes used in the experiment.

Presence of strong phonon modes in far- and mid-IR regions results in a higher contrast of 1.4 and 2, respectively, between Si and $SiO_2$ in white-light images collected using the corresponding experimental configurations, when compared to the THz near-field imaging (compare panels in Figure 3A).

The broadband complex spectral response of the $SiO_2$, referenced to Si, shows the spectral changes of the real and imaginary parts of the material dielectric function (Figure 3C and 3D). In particular, one can observe the characteristic phonon modes of Si-O-Si groups: rocking mode, as well as symmetric and asymmetric stretching modes at 500 $cm^{-1}$, 830 $cm^{-1}$ and 1030-1300 $cm^{-1}$,[27] respectively (Figure 3C and 3D). The experimental data agree well with the FDM simulated spectra obtained using the tabulated refractive index data of the material.

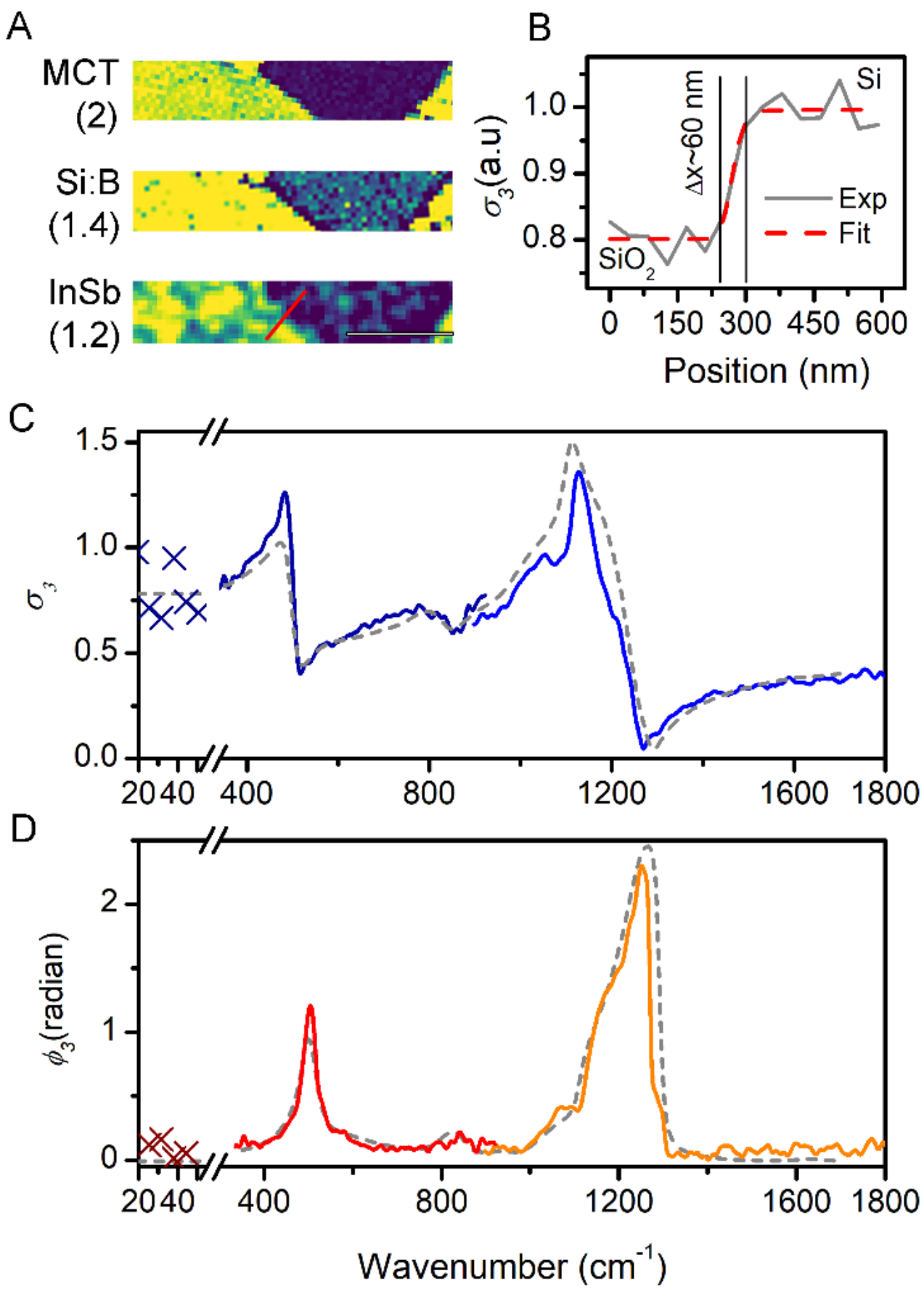

*Figure 3 A) Near-field optical amplitude images ($\sigma_3$) of the composite Si/$SiO_2$ sample using THz (InSb) light in the low-alpha mode and broadband synchrotron emission in mid-IR (MCT) and far-IR (Si:B) spectral ranges in multi-bunch operational mode of BESSY II storage ring. Dark blue and bright yellow areas correspond to $SiO_2$ and Si regions respectively. The values of 2, 1.4, and 1.2 correspond to the ratio of the optical amplitudes measured in Si and $SiO_2$ regions, for MCT, Si:B and InSb HEB detectors, respectively. B) Intensity profile of the near-field THz optical signal along the red line, as indicated in panel A of the figure. The estimated spatial resolution of the nano-IR technique in the low-alpha operational mode is 60 nm. C) and D) Near-field amplitude ($\sigma_3$) and phase ($\varphi_3$) spectra of $SiO_2$ obtained by concatenation of the THz, far- and mid-IR spectra collected with use of the three detectors. The solid lines and the cross symbols (X) represent the experimental data, the dashed traces are the FDM simulation spectra.*

In this work we demonstrate that the use of low-alpha operational mode of an electron storage ring provides significant brilliance benefit and enable scattering type near-field imaging and spectroscopy in the THz range at the energies down to 25 $cm^{-1}$ (*λ=0.4 mm*). It is shown that the system is capable to achieve AFM-probe defined spatial resolution in order of few tens of nanometers and detect the signal at higher demodulation harmonics, ensuring acquisition of the background-free near-field optical response. The unmodified interferometric design of the system and the broadband light source allow to perform the THz near-field spectroscopy. The use of the low-alpha operational mode allows the extension of the spectral range available at the synchrotron facilities, while the use of several optical detectors enable spectrally seamless s-SNOM based imaging and spectroscopy in the energy range above 25 $cm^{-1}$ covering more than 7.5 octaves of the electromagnetic spectrum.

**Acknowledgements**

We thank HZB-BESSY for the allocation of beam times at beamline IRIS. Funding by the German Federal Ministry for Education and Research (BMBF) project 05K19KH1 (SyMS) is gratefully acknowledged.

SUPPLEMENTARY MATERIAL

for

# THz near-field nanoscopy employing coherent synchrotron radiation

Alexander Veber[1,2], Maria Eleonora Temperini[1], Ljiljana Puskar[1], Ulrich Schade[1]

[1] *Institute for Electronic Structure Dynamics, Helmholtz-Zentrum Berlin für Materialien und Energie GmbH, Albert-Einstein-Str. 15, 12489 Berlin, Germany*

[2] *Department of Chemistry, Humboldt-Universität zu Berlin, Brook-Taylor-Straße 2, 12489 Berlin, Germany*

## Spectral response of the detector and the light flux of the coherent synchrotron radiation at BESSY II storage ring

Spectral responsivity of QFI/XBI InSb hot electron bolometer detector is reproduced from the operation manual supplied with the detector. The spectrum of the coherent synchrotron radiation at the IRIS beamline, BESSY II storage ring was measured using a superconducting transition edge Nb Bolometer (QMC Instruments Ltd, UK) coupled to a Bruker Vertex 70V spectrometer (Bruker Optik GmbH, Germany).

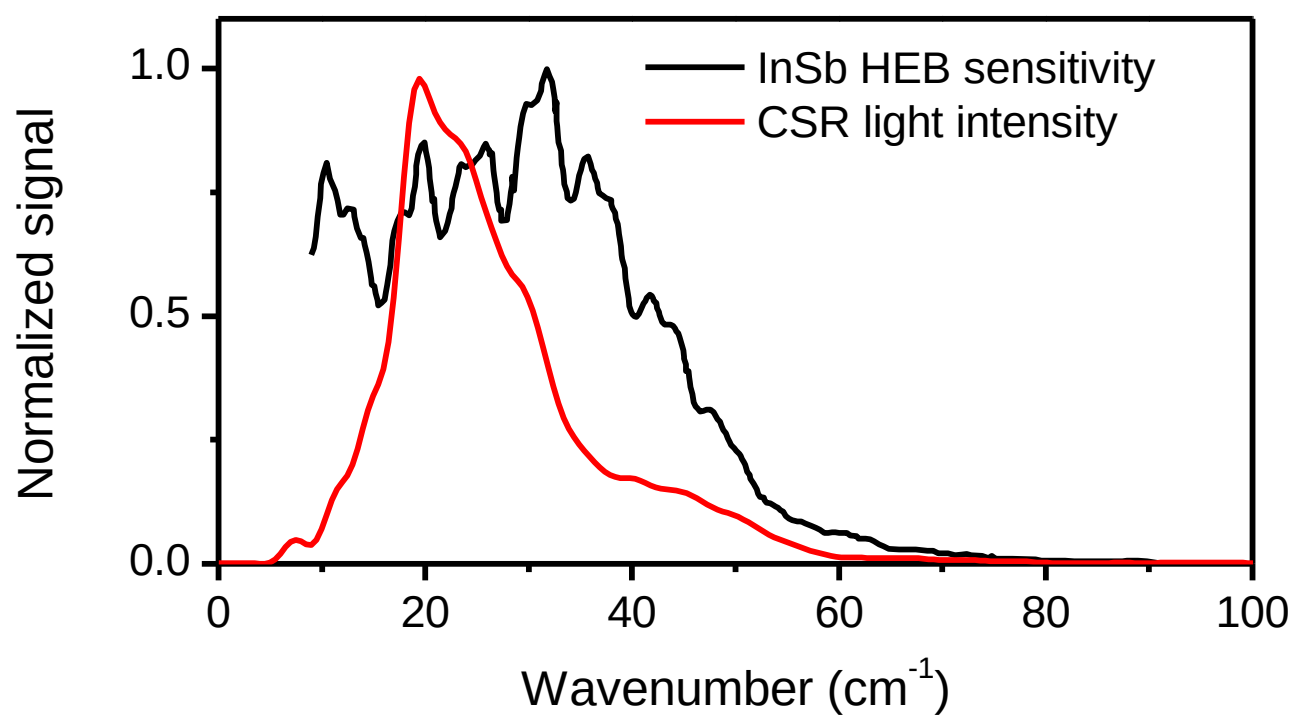

*Figure S1. Spectral responsivity of QFI/XBI InSb Hot electron bolometer detector (black trace) and intensity of the coherent synchrotron radiation at IRIS beamline BESSY II storage ring (red trace)*

**Optical configuration of the detector**

To suppress the incoming background radiation, we have installed a circular metallic aperture of approximately 8 mm before the long pass multi-mesh filter (cut-off edge is at 100 $cm^{-1}$) in the baffle of the detector at 77 K.

The light scattered from the AFM probe is collimated by the main parabola of the s-SNOM and then focused by an off-axis parabolic mirror with focal length of 10 cm at the installed aperture. The aperture size of 8 mm allows to accept the entire beam from the AFM-tip. The implemented detection outline, in particular the significantly reduced incoming aperture (from 25 mm to 8 mm) allows to match the optical étendue of the detector, thereby suppressing the incoming background radiation.

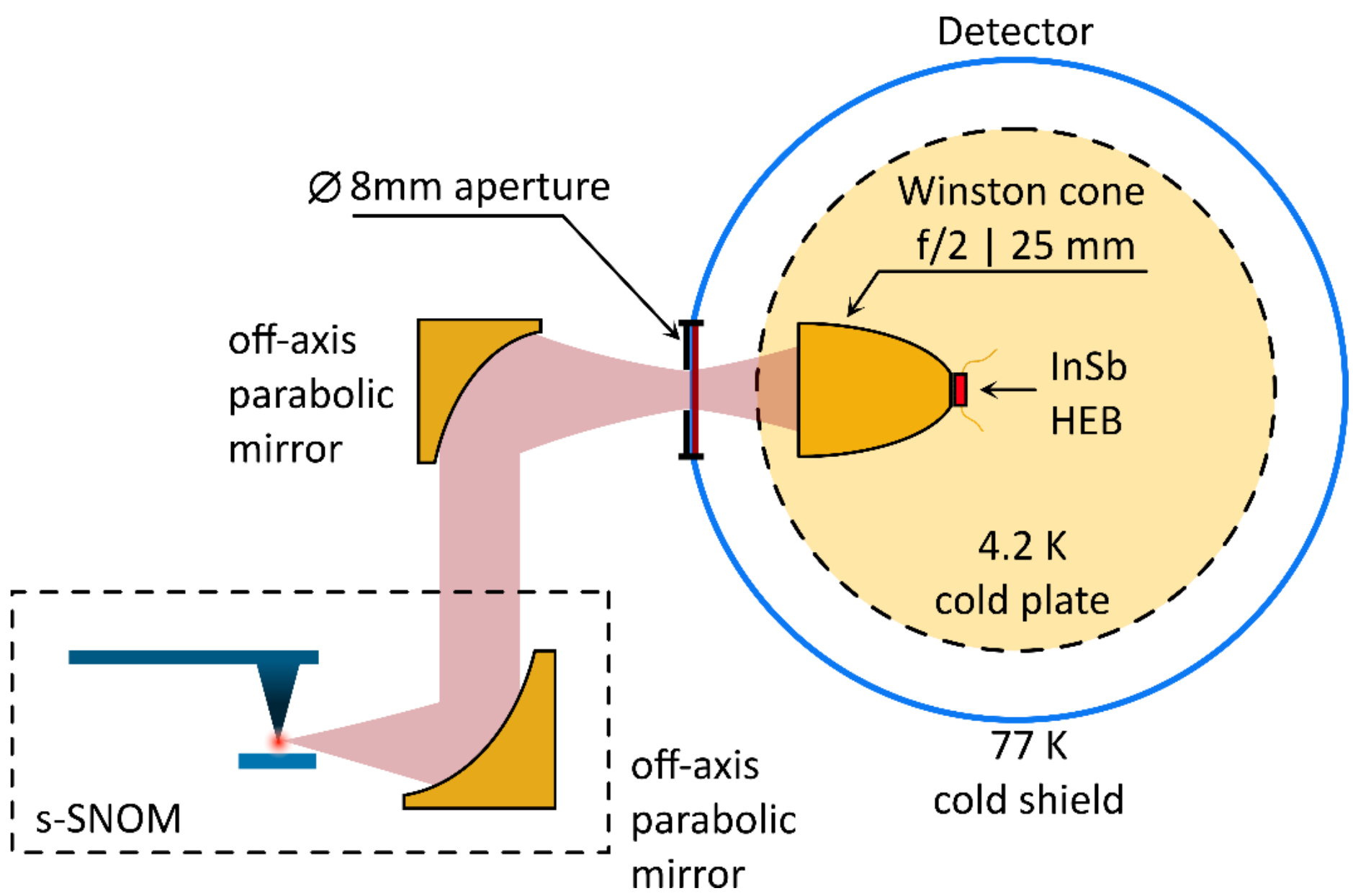

*Figure S2. Optical outline of the experimental setup. An additional aperture of ∅8 mm was installed in the baffle of the 77K cold shield before the 100 $cm^{-1}$ long pass filter. The light scattered from the s-SNOM AFM-probe is collimated by a parabolic mirror and then refocused at the detector's aperture.*